\documentclass{ws-procs961x669}            
\usepackage{lipsum}  

\usepackage{hyperref}
\begin{document}
\title{Probing Nuclear Structure with Future Colliders\footnote{Contribution to
The Institute for Nuclear Theory Program INT-18-3: ``Probing Nucleons and Nuclei in High Energy Collisions." October 1 - November 16, 2018}}

\author{
T. J. Hobbs\rlap,${}^{a,b}$ 
Pavel~M. Nadolsky\rlap,${}^{a}$ 
Fredrick I. Olness\rlap,${}^{a}$\footnote{Presenter.}
Bo-Ting Wang${}^{a}$ 
}
\address{${}^{a}$Department of Physics, Southern Methodist University,\\
 Dallas, TX 75275-0175, U.S.A. }
 \vspace{-0.4cm}
\address{${}^{b}$Jefferson Lab, EIC Center, Newport News, VA 23606, U.S.A.}


\begin{abstract}

Improved knowledge of the nucleon structure is a crucial pathway toward a deeper understanding 
of the fundamental nature of the QCD interaction, and will enable important future discoveries. 
The  experimental facilities proposed for the next decade offer a tremendous opportunity
to advance the precision of our theoretical predictions to unprecedented levels.  
In this report we briefly highlight some of the recently developed tools and techniques 
which, together with data from these new colliders, have the potential to revolutionize our understanding of the QCD theory in the next decade.
\end{abstract}
%

\bodymatter

\def\figone{
\begin{figure*}[t]
\centering
\includegraphics[clip,width=1.00\textwidth]{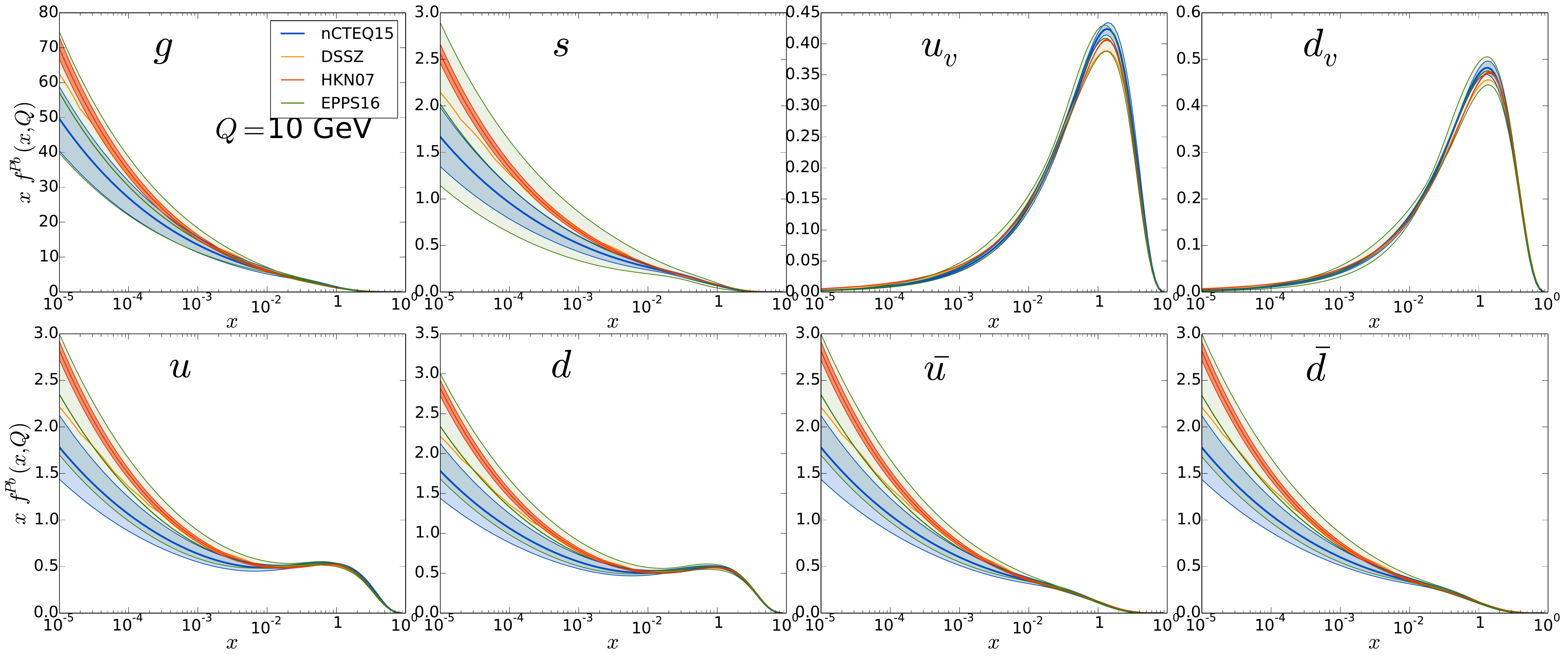}
	\caption{
	Comparison of selected nuclear PDF flavors and uncertainties including 
	HKN07, DSSZ,
	nCTEQ15\cite{Kovarik:2015cma}
	and
	EPPS16.\cite{Eskola:2016oht}
	}
\label{fig:npdf}
\end{figure*}
}

\def\figtwo{
\begin{figure*}[th]
\includegraphics[clip,width=0.66\textwidth]{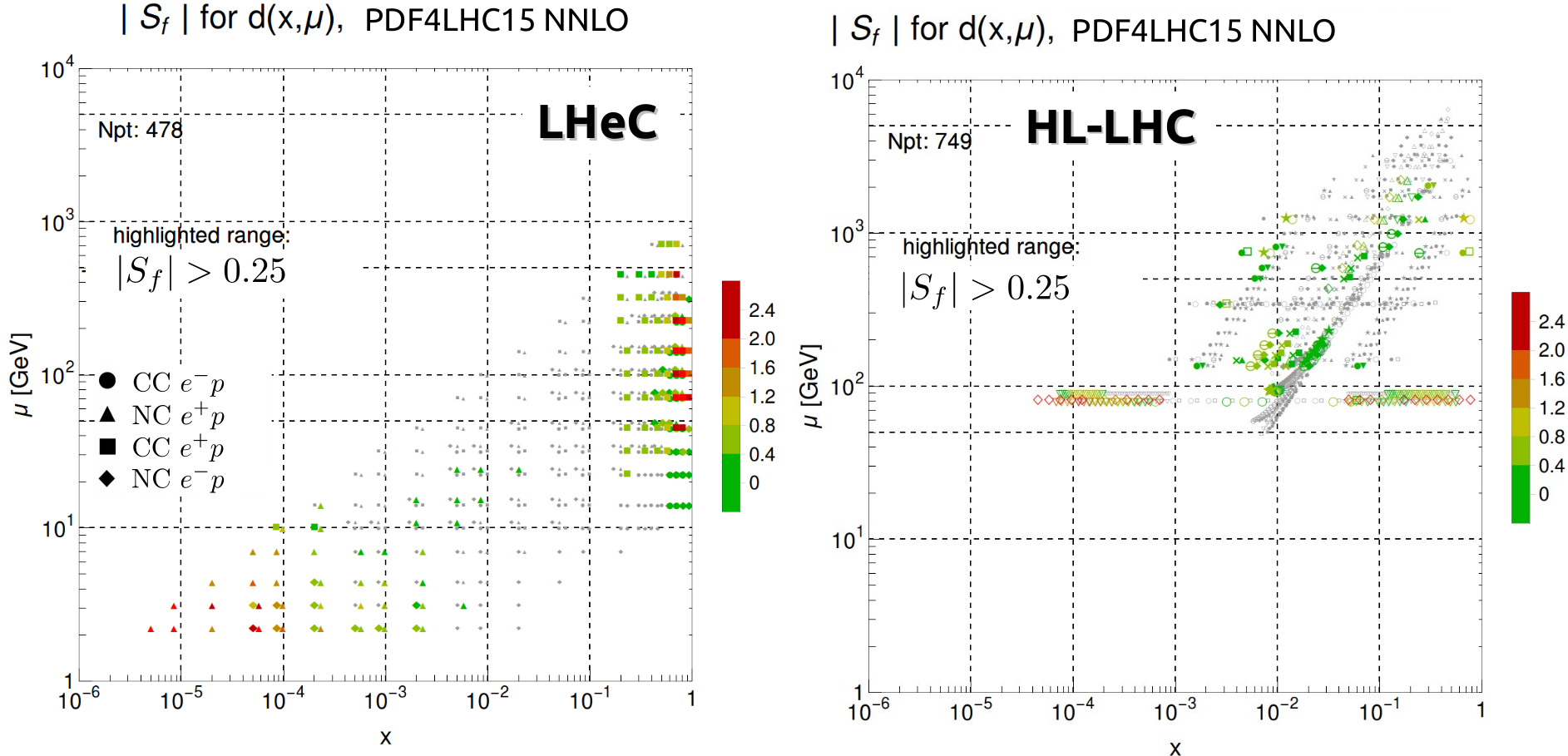}
\hfil 
\includegraphics[clip,width=0.33\textwidth]{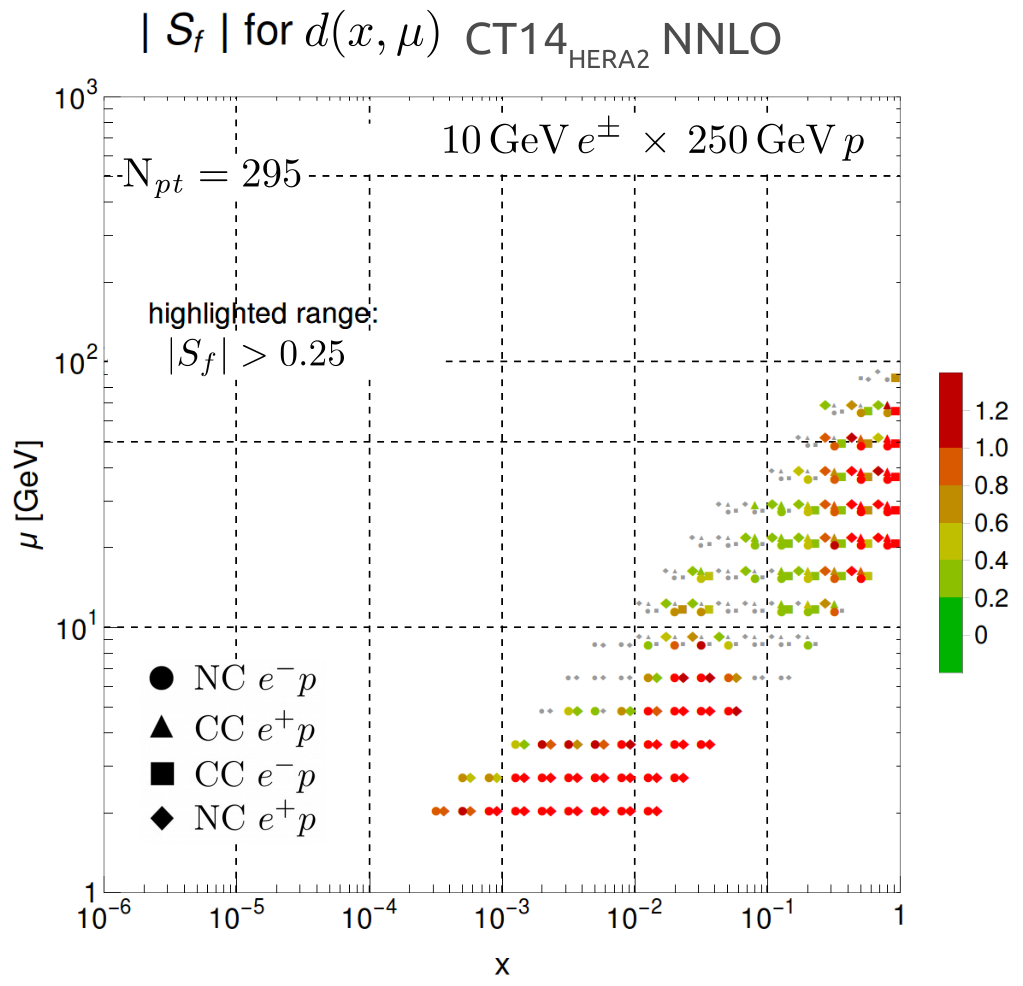}
	\caption{
	Future HEP experiments such as the  LHeC, HL-LHC, and EIC  can have substantial PDF
	sensitivity $S_f$ as shown here for the   $d(x,\mu)$ distribution computed according to the
	conventions in Refs.~\citenum{Wang:2018heo}, and~\citenum{Hobbs:2019sut}. 
	}
\label{fig:sens}
\end{figure*}
}

\def\figthree{
\begin{figure}[t]
\begin{minipage}{0.32\textwidth}
\centering
\includegraphics[clip,width=0.99\textwidth]{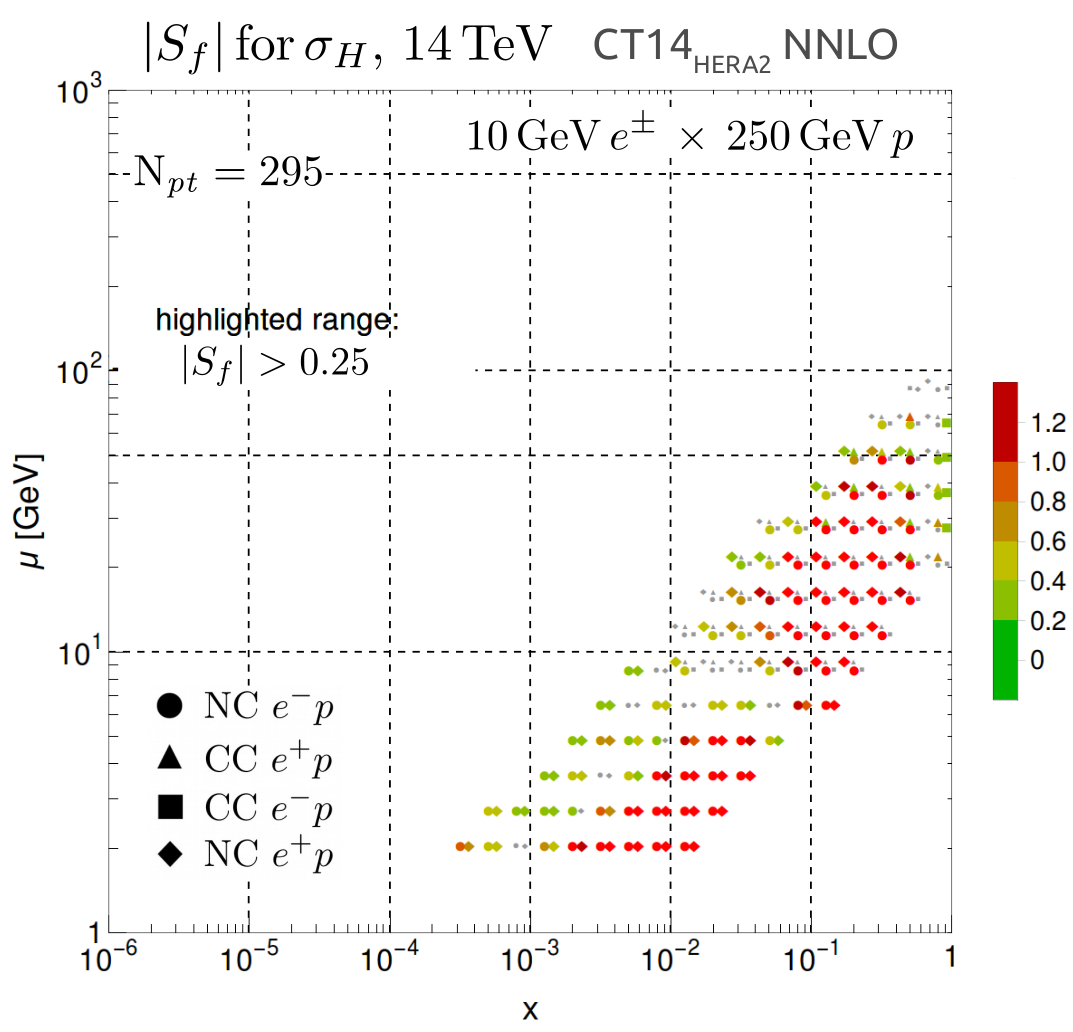}
	\caption{The EIC  pseudodata
sensitivity  for Higgs
production\cite{Hobbs:2019sut} with  $\int L = 100~fb^{-1}$.
}
\label{fig:higgs}
\end{minipage}
\hfil
\begin{minipage}{0.32\textwidth} 
\centering
\includegraphics[clip,width=0.99\textwidth]{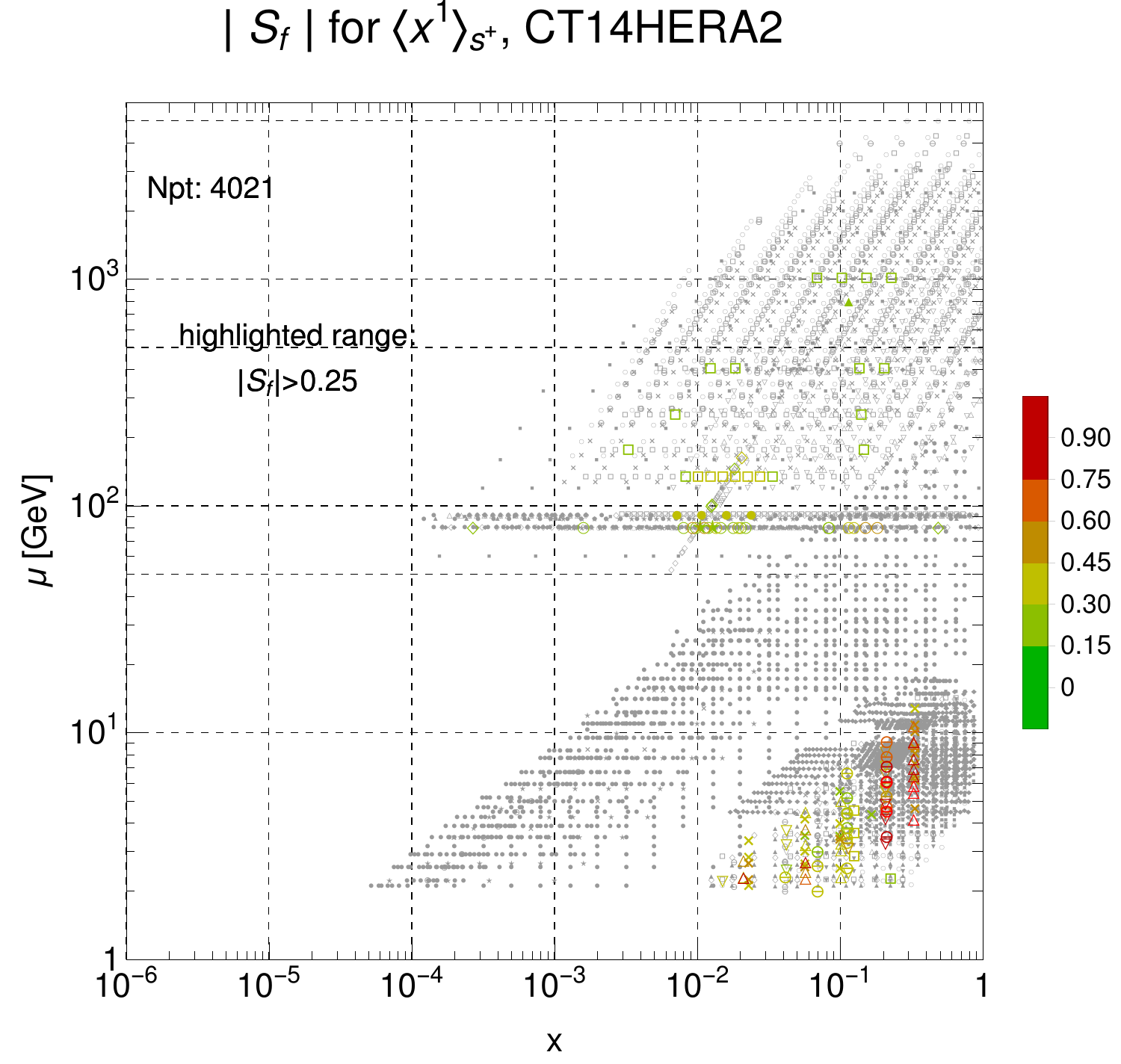}
	\caption{Sensitivity to the 1st-order Mellin moments of the $s^+$ distribution;\cite{Hobbs:2019gob} $\mu_F=2$~GeV.
	}
\label{fig:moment}
\end{minipage}
\hfil
\begin{minipage}{0.32\textwidth}
\centering
\includegraphics[clip,width=0.99\textwidth]{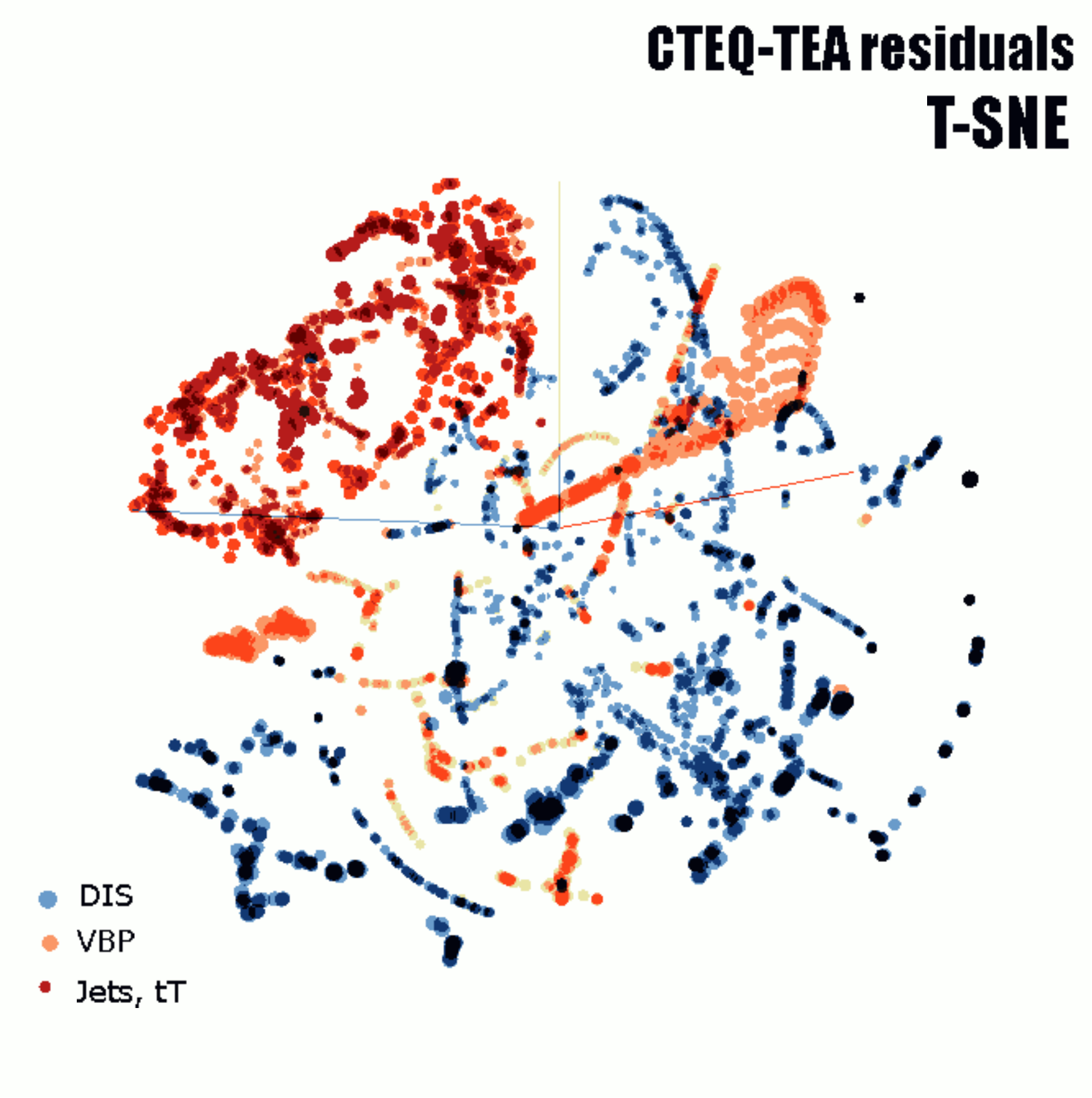}
	\caption{
 Distribution from the t-SNE
clustering method.\cite{Wang:2018heo}  
\vspace{20pt}
	}
\label{fig:tsne}
\end{minipage}
\label{fig:HIGGS}
\end{figure}
}

\def\head#1{\vspace{0.2cm}\noindent{\textbf{#1:}}}

\vspace{0.5cm}
\head{Introduction}
\label{sec:intro}
\hyphenpenalty=1000
The recent discovery of the Higgs boson at the Large Hadron Collider (LHC) was a remarkable endeavour which culminated in the  2013 Nobel Prize.
As we look ahead to the next decade,   a number of new facilities on 
the horizon 
will provide new insights about fundamental microscopic phenomena.
This includes the high-luminosity upgrade of the LHC, (HL-LHC), 
a proposed electron ring for the LHC (LHeC),\footnote{A Large Hadron electron Collider at CERN
\href{http://lhec.web.cern.ch/}{http://lhec.web.cern.ch/}
} and a proposed Electron-Ion Collider (EIC).\footnote{Electron-Ion Collider (EIC) User Group
\href{http://www.eicug.org/}{http://www.eicug.org/}}

For all these facilities, the route to new discoveries will be  high-precision comparisons between theory and data that can validate the features of the Standard Model and search for discrepancies which may signal undiscovered phenomena. 

These comparisons between data and theory rely crucially on the Parton Distribution Functions (PDFs) which connect the theoretical quarks and gluons with experimental observations.  Unfortunately, the PDFs are often the limiting factor in these comparisons.
Thus, our ability to fully characterize the Higgs boson and constrain physics Beyond the Standard Model (BSM) ultimately comes down to how accurately we determine the underlying PDFs. If you cannot distinguish signal from background, you cannot make discoveries. 

In this brief report, we will examine how new data and new tools might facilitate these discoveries.\footnote{The articles cited here are limited due to space; please also see references therein.}

\figone

\head{Nuclear PDFs}
Although the fits to the proton PDFs have been quite successful, 
it is crucial to extend the framework to consider the nuclear degrees of freedom. For example, the neutrino-induced Deeply Inelastic Scattering (DIS) process provides essential information for PDF flavor differentiation. 
In Fig.~\ref{fig:npdf} we display a comparison of recent nuclear PDFs with uncertainties. While there has been impressive progress in recent years 
constraining the nPDFs, there is certainly room for improvement, especially compared to the proton PDF efforts; the new experimental facilities 
(including LBNF)
will provide important new PDF constraints which in turn will constrain 
potential BSM signals. 


\head{PDF Sensitivity \& PDFSense\cite{Wang:2018heo,Hobbs:2019gob}}
In order to quickly and efficiently determine the potential impact of 
new data sets on the PDFs and other observables,  we introduce a generalization of the PDF-mediated correlations called the sensitivity $S_f$.
This is a combination of the correlation coefficient and the residuals (data minus theory scaled by the uncertainty) of the PDF uncertainty,
and it  identifies those experimental data points that tightly 
constrain the PDFs. 
We find the sensitivity is useful for identifying regions of the $\{x,Q^2\}$ kinematic plane in which the
PDFs are particularly constrained by physical observables.
The details of the sensitivity can be found in Ref.~\citenum{Wang:2018heo} and the implementation is available in the public package
PDFSense.\footnote{\href{https://metapdf.hepforge.org/PDFSense/}{https://metapdf.hepforge.org/PDFSense/}}

\figtwo

In Fig.~\ref{fig:sens} we display the sensitivities $S_f$ for
the LHeC, HL-LHC, and EIC using pseudodata sets for a sample PDF flavor, $d(x,Q)$, {\it c.f.,} Ref.~\citenum{Hobbs:2019sut}.
We observe the LHeC shows strong sensitivity in both the very high- and low-$x$ regions, while the HL-LHC covers the intermediate-$x$ region out to very large $Q^2$, and the  EIC complements these in the high-to-medium-$x$ region at 
lower $Q^2$ values. 
The combination of these measurements can provide very strong constraints on
the various PDF flavors across the broad  $\{x,Q^2\}$ kinematic plane.


\newpage
\null \vspace{-18pt}
In Fig.~\ref{fig:higgs}, we plot the sensitivity of EIC pseudodata to the
14 TeV LHC  Higgs production cross section, $\sigma_H$.
The constraints that a medium-energy machine like an EIC would place
on Higgs phenomenology stem from the predominance of the $gg \to  H$
fusion channel at the LHC, and the 
sensitivity of the DIS data to the gluon via QCD evolution, which  
then connects the high-$x$ and low-$Q$ gluon PDF probed by the EIC to the lower-$x$ and high-$Q$ PDF of the LHC.
The sensitivity of the EIC pseudodata generally surpasses that of
the fixed-target experiments that currently dominate the constraints
on high-$x$ PDFs.
Therefore, the EIC will  strongly constrain the PDF dependence of HEP
observables at moderate to  large-$x$, including several in the Higgs
and electroweak sectors, like $M_W$ and $\sin (2 \theta_W)$.


\head{PDFs from lattice QCD}
Additional information on PDFs may also come from lattice QCD
calculations\cite{Lin:2017snn} which generally compute the Mellin moments of PDFs, as illustrated in
Fig.~\ref{fig:moment}, where we display the sensitivity, $S_f$, of the first moment for the
strange quark.\cite{Hobbs:2019gob}
Despite
additional data from HERA and LHC, 
it is interesting to note that some of the strongest
sensitivities come from the fixed-target neutrino DIS experiments
(CCFR, NuTeV) in the high-$x$ and low-$Q$ region.  
At present, the strange PDF still has relatively
large PDF uncertainties, and this can limit the precision of ``standard candle" benchmark
process such as $W/Z$ production; hence, additional constraints from
lattice QCD would be welcome.

\head{Borrowing from Machine Learning}
Finally, in  Fig.~\ref{fig:tsne} we display a  distribution of 
residuals using a
t-distributed Stochastic Neighbor Embedding (t-SNE) which 
is a machine-learning algorithm for visualization of high-dimensional data.
In this case, we have taken the 56-dimensions of the PDF uncertainty eigenvectors and reduced this to a 3D projection of the 4000+ data points considered for the CT18 fit. The algorithm identifies points with similar characteristics and 
groups these together. We have color coded the points for the DIS, 
Drell-Yan, and Jets+$t\bar{t}$ sets.
In this case, the fact that the algorithm has grouped similar experiments together suggests that such machine-learning techniques can help us identify subtle relations that may not be apparent via other methods.\cite{Wang:2018heo}

\figthree

\goodbreak
\newpage
\head{Conclusions}
We have highlighted a few of the tools and techniques that will allow us to most effectively 
capitalize on the wealth of new data from these proposed colliders to advance our knowledge
of the structure of the nucleon, the associated nuclear PDFs, and the underlying QCD theory. 
Additionally, these tools demonstrate how 
each of these experimental facilities occupies a unique place in the kinematical parameter space. 
As such, these complementary data sets will provide us with unprecedented understanding of  strong interaction physics, 
and provide the key for many new discoveries.


\head{Acknowledgments}
We are grateful to our colleagues of the CTEQ-TEA, nCTEQ, and xFitter collaborations.
We also thank the Institute for Nuclear Theory at the University of Washington for its kind hospitality and stimulating research environment.
This work was supported by the U.S.\ Department of Energy under Grant No.\ {D}{E}-SC0010129
and DE-FG02-00ER41132.
The research of TJH is supported by an EIC Center@JLab Fellowship.

\bibliography{./Biblio/main,./Biblio/extra}
\bibliographystyle{ws-procs961x669}

\end{document}